\documentclass{webofc}
\usepackage[english]{babel}
\usepackage[utf8]{inputenc}
\usepackage{amsmath,ulem}
\usepackage{amssymb}
\usepackage{graphicx}
\usepackage{color}
\usepackage{float}
\usepackage{slashed}
\usepackage{lipsum}
\usepackage[varg]{txfonts}   % Web of Conferences font

\begin{document}
\title{Hybrid stars with large strange quark cores}

        \author{\firstname{Márcio Ferreira}\inst{1}\fnsep\thanks{\email{marcio.ferreira@uc.pt}} 
        \and \firstname{Renan Câmara Pereira} \inst{1}\fnsep\thanks{\email{renan.pereira@student.uc.pt}}
        \and \firstname{Constança Providência} \inst{1}\fnsep\thanks{\email{cp@uc.pt}}}

\institute{CFisUC, 
	Department of Physics, University of Coimbra, P-3004 - 516  Coimbra, Portugal}

\abstract{%
The possible existence of hybrid stars is studied using several multi-quark interaction channels. The hadronic phase consists of an equation of state (EoS) with presently accepted nuclear matter properties and the quark model is constrained by the vacuum properties of several light mesons. The dependence of several NS properties on the different quark interactions is analyzed. We show that the present constraints from neutron star observations allow for the existence of hybrid stars with large strangeness content and large quark cores.
}
\maketitle

\section{Introduction}
\label{intro}
The association of astrophysical observations from
electromagnetic radiation, astro-particles, and gravitational waves (GW) -- multi-messenger astrophysics - allows to explore the physics of neutron stars (NSs) and draw some information concerning the high density behavior of the QCD phse diagram. The observation of massive pulsars, PSR J0348+0432 with $M=2.01\pm$0.04 $M_\odot$
\cite{Antoniadis:2013pzd} and MSP J0740$+$6620 with $2.08^{+0.07}_{-0.07}M_\odot$ 
\cite{Cromartie2019,Fonseca2021}, sets strong constraints on the properties of neutron star (NS) matter. From the the analysis of the GW170817 event, by the LIGO/Virgo collaboration, it was possible to estimate an upper bond on the tidal deformability of a NS star, imposing additional constraints to the NS EoS \cite{TheLIGOScientific:2017qsa,LIGOScientific:2018cki}.
While the inferred upper bond on the tidal deformability from the GW170817 event favors not too stiff EoS, massive pulsars require the opposite behavior, i.e., a considerably stiff EoS. %Notice, however, that the GW190425 seems to indicate larger values of the tidal deformability \cite{GW190425}.
This tension might indicate the possible existence of exotic matter, such as deconfined quark matter, inside NSs. The possible existence of a quark core in a $1.4M_{\odot}$ NS would give rise to smaller tidal deformabilities, consistent with the GW170817.

This work explores precisely this scenario, i.e., the existence of a quark core at moderate NS masses and still consistent with massive NSs. We describe the phase transition from hadronic to quark matter 
using a Maxwell construction and explore NJL-type models \cite{Ferreira:2020evu,Ferreira:2020kvu,Ferreira:2021osk} in order to reproduce not only quark cores at moderate NS masses but also massive NSs. 

\section{Equation of state of hybrid NS}
\label{eos}
We describe hybrid stars using a two-model approach, where the hadronic and quark phases are realized via a first-order phase transition.
We have selected the DDME2 \cite{ddme2} for the hadronic phase, which has good nuclear matter properties at saturation: $K_{sat}=250$~MeV, $E_{sym}=32.3$~MeV, $L_{sym}=51.2$~MeV, and $K_{sym}= -87$~MeV. 
The quark phase is described by the following three-flavor NJL-type model 
\begin{align}
\mathcal{L} &=  
\bar{\psi} 
(
i\slashed{\partial} - \hat{m} + \hat{\mu} \gamma^0 
) 
\psi 
 + G_S  \sum_{a=0}^8
[ (\bar{\psi} \lambda^a \psi)^2 + 
(\bar{\psi} i \gamma^5 \lambda^a \psi)^2 ] \nonumber
\\
 &- G_D [  
\det( \bar{\psi} (1+\gamma_5) \psi ) + 
\det( \bar{\psi} (1-\gamma_5) \psi )  ] \nonumber
\\
&  - G_\rho \sum_{a=1}^8 
[ 
(\bar{\psi} \gamma^\mu\lambda^a \psi)^2 +  
(\bar{\psi} \gamma^\mu\gamma_5\lambda^a \psi)^2 
]
- G_{\omega \omega} 
[ (\bar{\psi}\gamma^\mu\lambda^0\psi)^2 + (\bar{\psi}\gamma^\mu\gamma_5\lambda^0\psi)^2 ]^2
\end{align} 
The model parameters, $\hat m=diag(m_u,m_d,m_s)$, $G_S$, $G_D$, and $\Lambda$, are fitted to reproduce the vacuum properties (see Table \ref{tab:2}). The effect of the (free) couplings $G_{\rho}$ and $G_{\omega \omega}$ on the neutron star properties are analyzed in the present work. The pressure and energy density for cold ($T=0$) quark matter is obtained from the thermodynamic potential, calculated using the  mean-field approximation.

\begin{table}[!htb]
\caption{Quark model parametrization. $\Lambda$ represents the momentum cutoff, $m_{u,d,s}$ are the quark current masses, $G_S$ and $G_D$ are coupling constants. Furthermore, the vacuum constituent quark masses, $M_{u,d}$ and $M_{s}$, are also shown. }
\begin{center}
    \begin{tabular}{cccccccc}
    \hline
    \hline
$\Lambda$  & $m_{u,d}$ & $m_s$  & $G_S\Lambda^2 $ & $G_D\Lambda^5 $ & $M_{u,d}$ & $M_s$ \\
\text{[MeV]}      &  [MeV]    & [MeV]   &               &                  & [MeV] &  [MeV]\\
\hline
   \hline
623.58 & 5.70   & 136.60 & 1.67 &  13.67 & 332.2   & 510.7 \\
   \hline
\end{tabular}
\label{tab:2}
\end{center}
\end{table}
The quark pressure $P$ and energy density $\epsilon$ are defined up to the bag constant $B$: $P \to P + B$ and $\epsilon \to \epsilon-B$. The value of $B$ allows to control the baryonic density at which the first-order phase transition between hadronic (H) to quark (Q) matter occurs. The phase transition is obtained by imposing the Maxwell construction: $\mu_B^H = \mu_B^Q$ and $P^H = P^Q$, where $\mu_B^{H,Q}$ is the baryon chemical potential of each phase. The next section analyzes the impact of the coupling ratios $\chi_{\rho} = G_\rho / G_S$, $\chi_{\omega \omega} = G_{\omega \omega} / G_S^4$ on the hybrid stars properties. 

\section{Results and discussion}
\label{sec:results}
\begin{figure}[!htb]
\begin{center}
\begin{tabular}{c}
\includegraphics[width=0.33\linewidth]{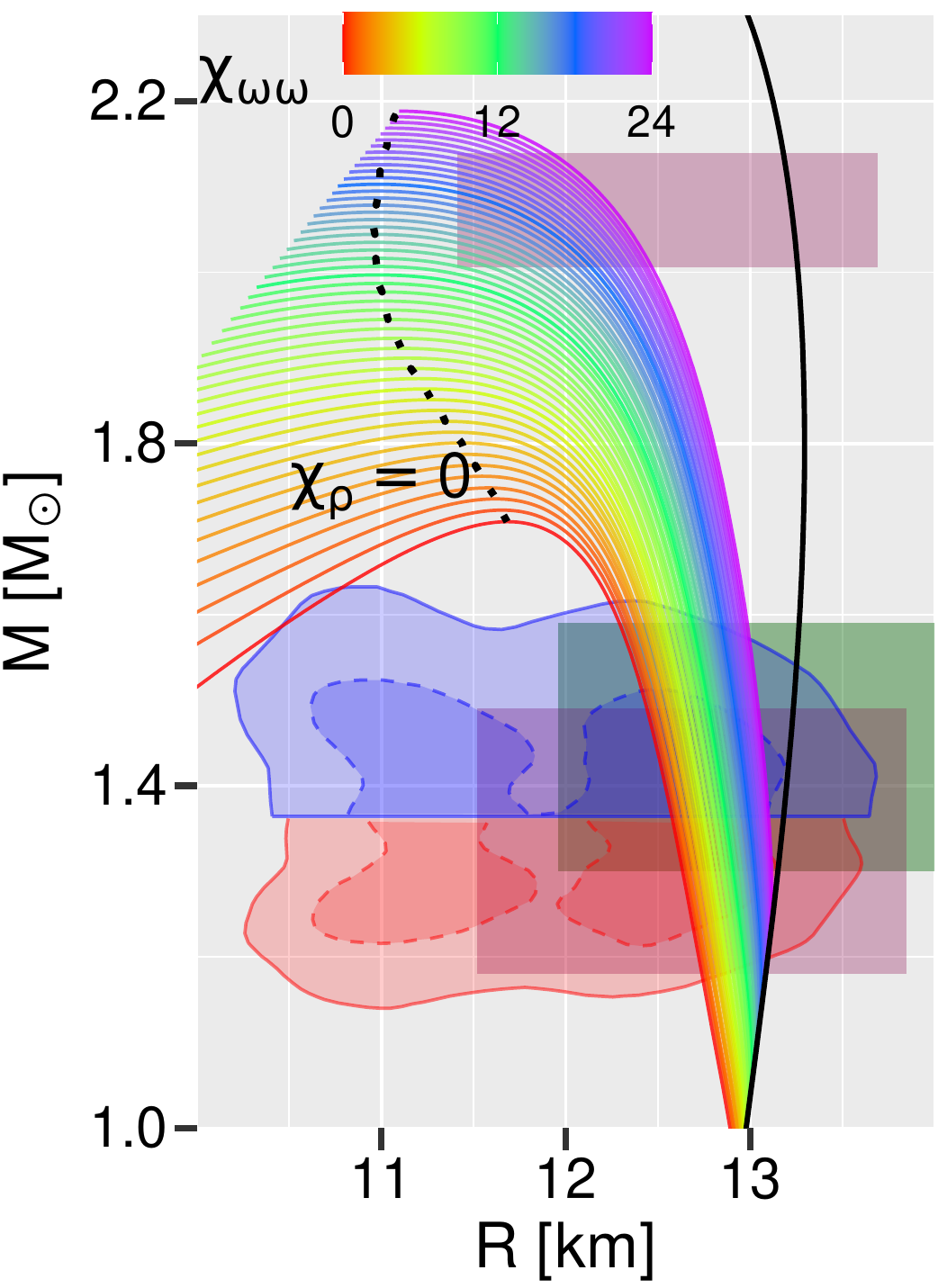}
\includegraphics[width=0.33\linewidth]{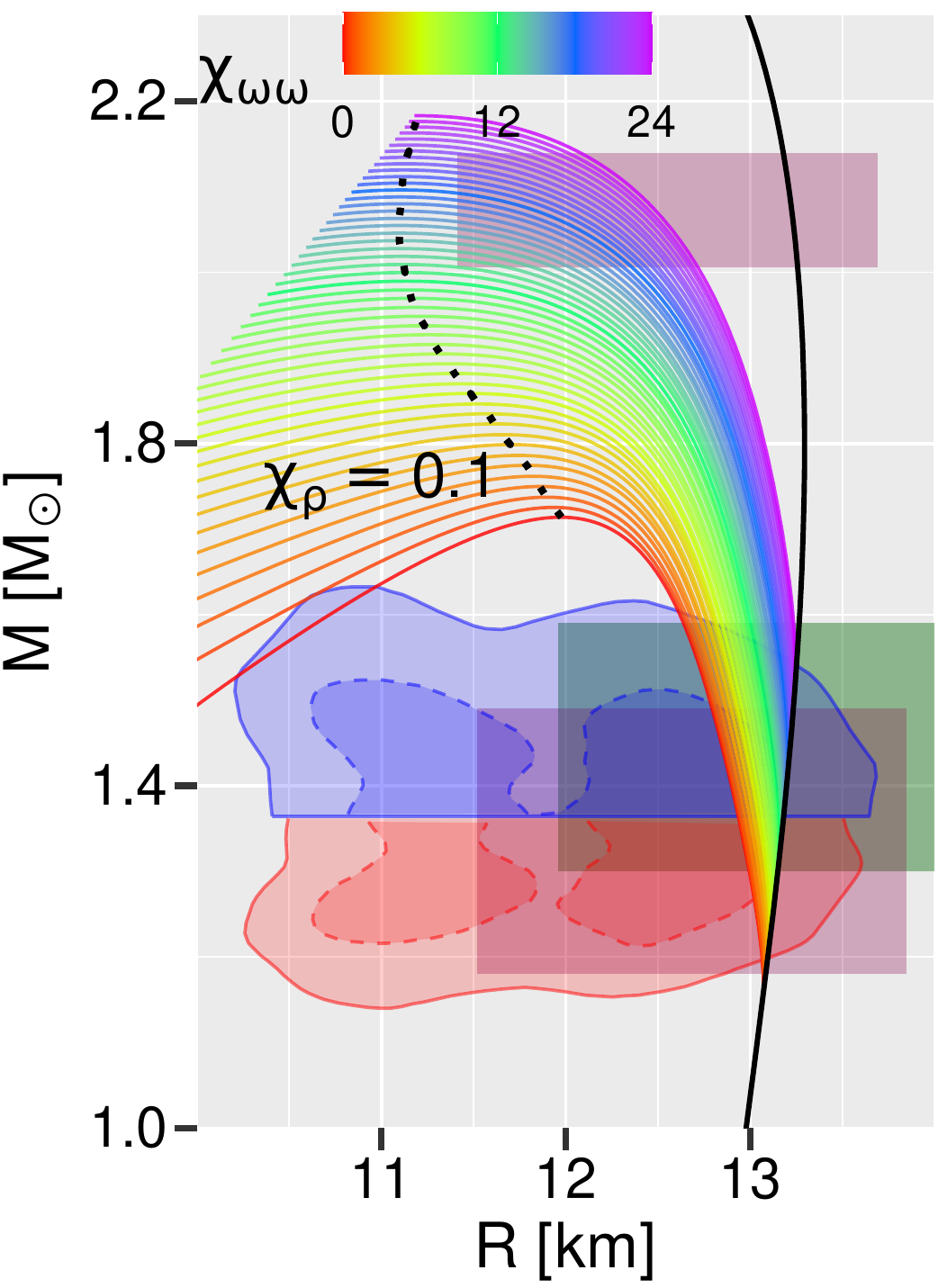}
\includegraphics[width=0.33\linewidth]{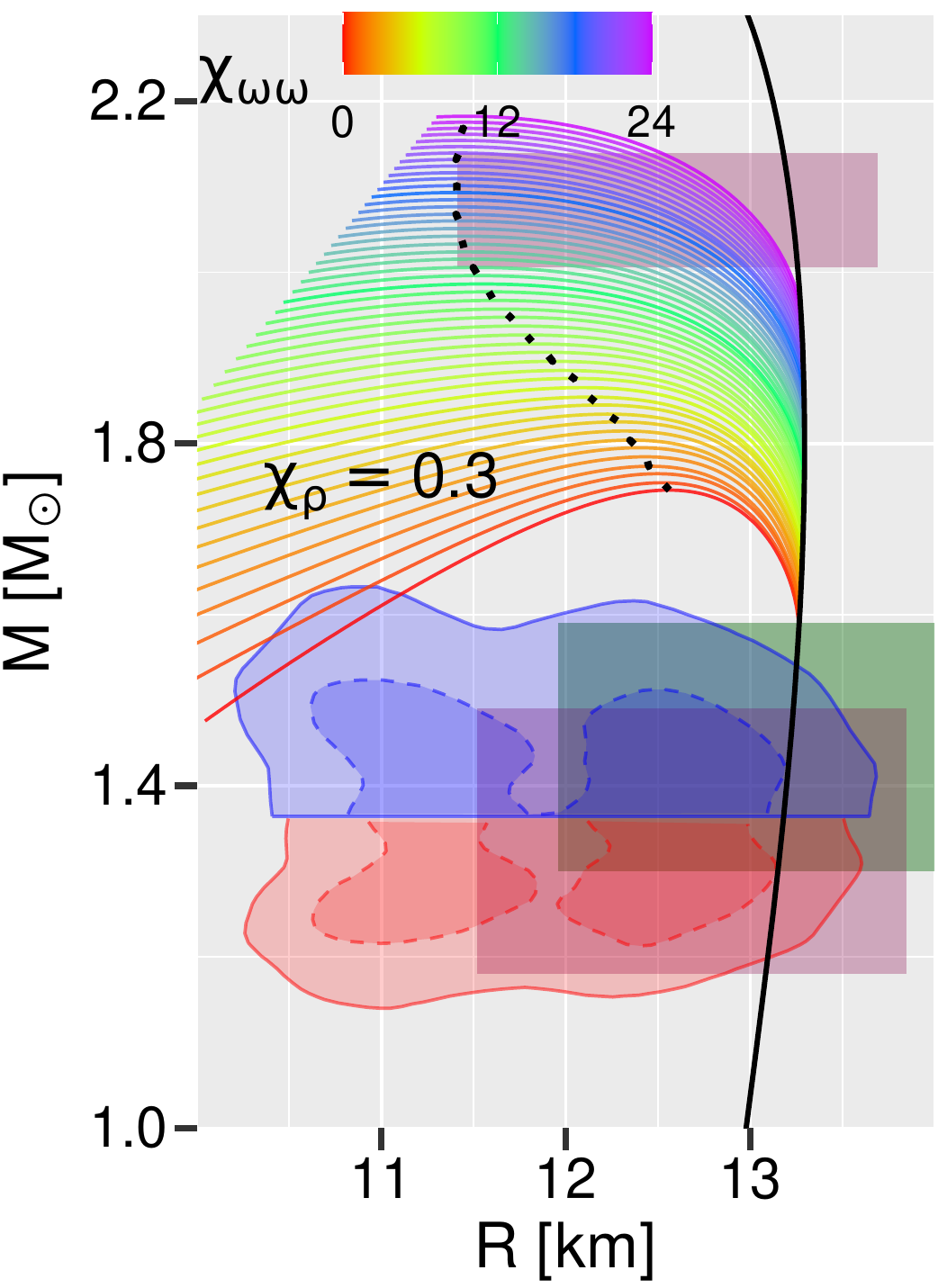}
\end{tabular}
	\caption{$M(R)$ diagrams as a function of $\chi_{\omega\omega}$ (color scale)  for three values of $\chi_{\rho}$: 0 (left), 0.1 (center), and 0.3 (right). The solid black line represents the purely hadronic $M(R)$ sequence, and the black dotted specifies the maximum mass of each hybrid EoS. The rectangular regions enclosed by dotted lines indicate the constraints from the millisecond pulsar PSR J0030+0451 NICER x-ray data  \cite{Riley:2019yda,Miller:2019cac}, while the top brown region is from \cite{Miller2021}. The top blue and bottom red regions indicate, respectively, the 90\% (solid) and 50\% (dashed) credible intervals of the LIGO/Virgo analysis for the GW170817 event \cite{LIGOScientific:2018cki}.}
	\label{fig:MR}
	\end{center}
\end{figure}

\begin{figure}[!htb]
\begin{center}
\begin{tabular}{c}
\includegraphics[width=0.33\linewidth]{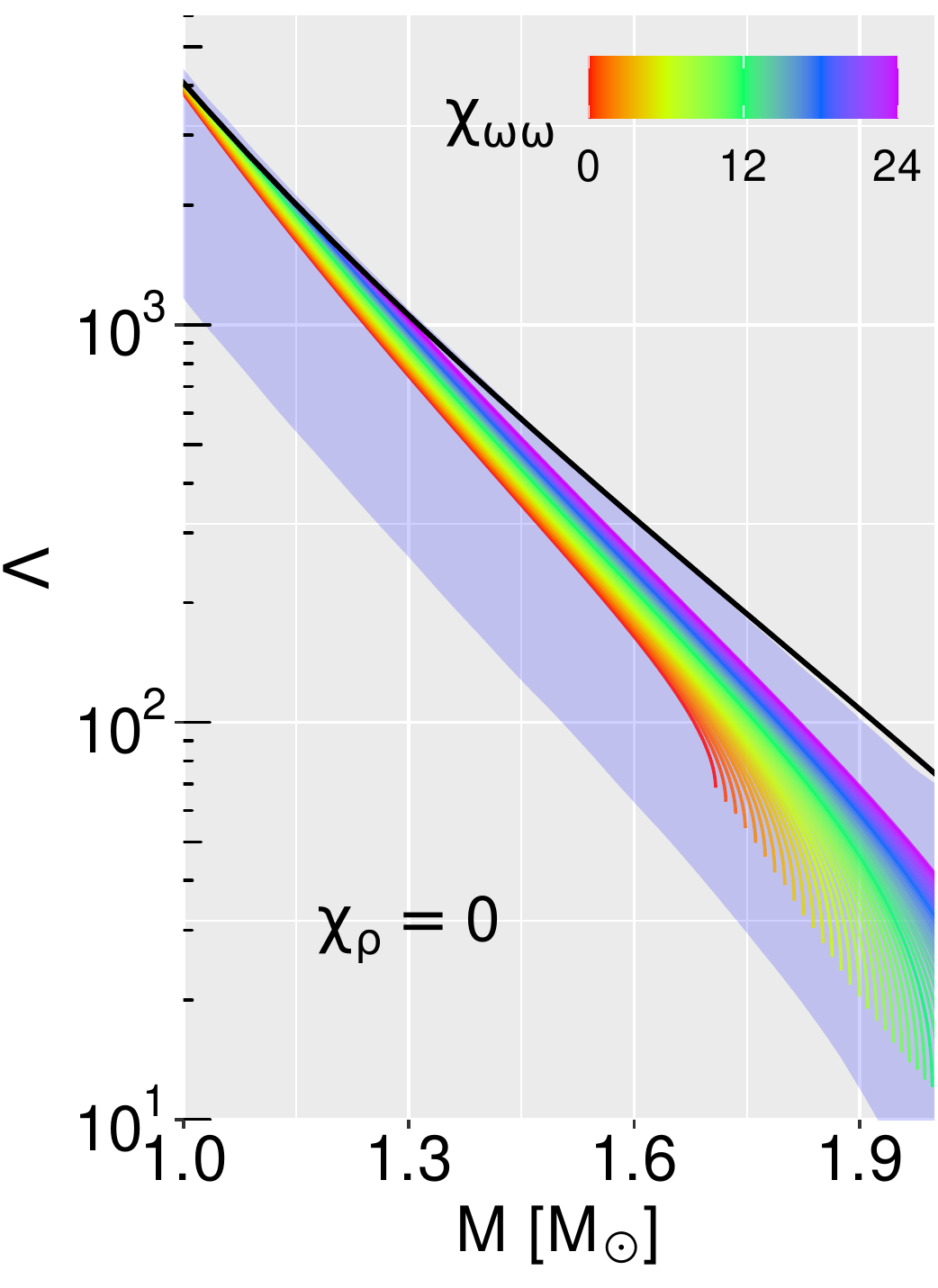}
\includegraphics[width=0.33\linewidth]{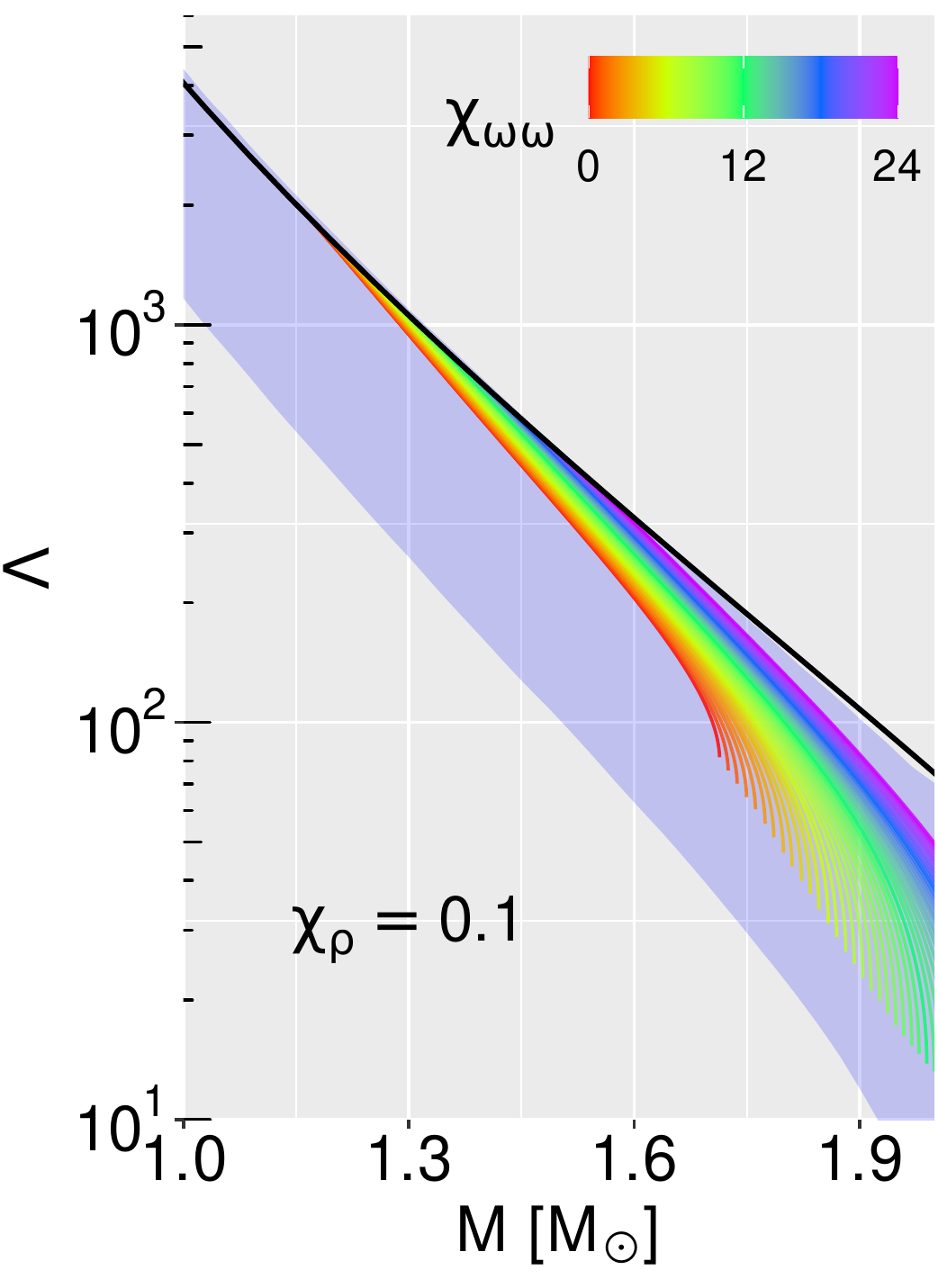}
\includegraphics[width=0.33\linewidth]{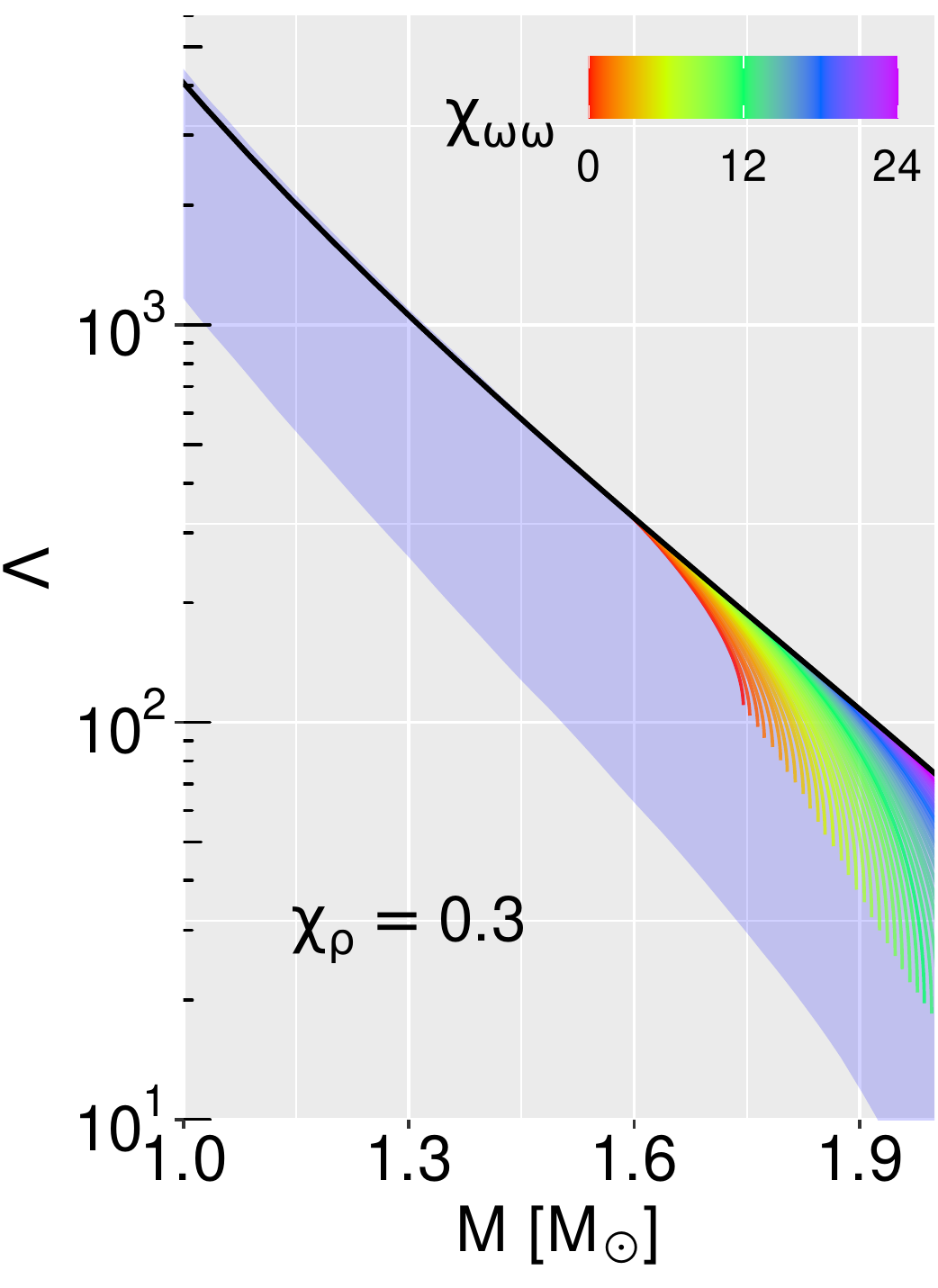}
\end{tabular}
	\caption{$\Lambda(M)$ diagrams as a function of $\chi_{\omega\omega}$ (color scale)  for three values of $\chi_{\rho}$: 0 (left), 0.1 (center), and 0.3 (right). The solid black line represents the purely hadronic $\Lambda(M)$ sequence, and the black dotted specifies the maximum mass of each hybrid EoS.  The blue band indicates the 90\%  posterior credible level from the LIGO/Virgo analysis for the GW170817 event \cite{LIGOScientific:2018cki}.}
	\label{fig:lambda}
	\end{center}
\end{figure}

The effect of $\chi_{\rho}$ and $\chi_{\omega \omega}$ on the $M(R)$ sequence is shown in Fig.~\ref{fig:MR}. As the value $\chi_{\rho}$ increases, from 0 (left) to 0.3 (right), the onset of the quark phase happens at larger NS masses, giving rise to shorter hybrid star branches. The effect of the repulsive interaction  $\chi_{\omega \omega}$ is clear: the larger the coupling strength the  heavier the hybrid NS; however,  the $\chi_{\omega \omega}$ also shifts the onset of quarks to higher NS masses.  
We see that there is a considerable set of hybrid star models $\{ \chi_{\rho}, \chi_{\omega \omega} \}$, with low $\chi_{\rho}$ values, compatible with the most recent NS observations. 
Furthermore, Fig. \ref{fig:lambda} indicates that the hybrid stars models are also compatible with the credible intervals extracted by the LIGO/Virgo collaboration analysis from the GW170817 event \cite{LIGOScientific:2018cki}. 

\begin{figure}[!htb]
\begin{center}
\begin{tabular}{c}
\includegraphics[width=0.95\linewidth]{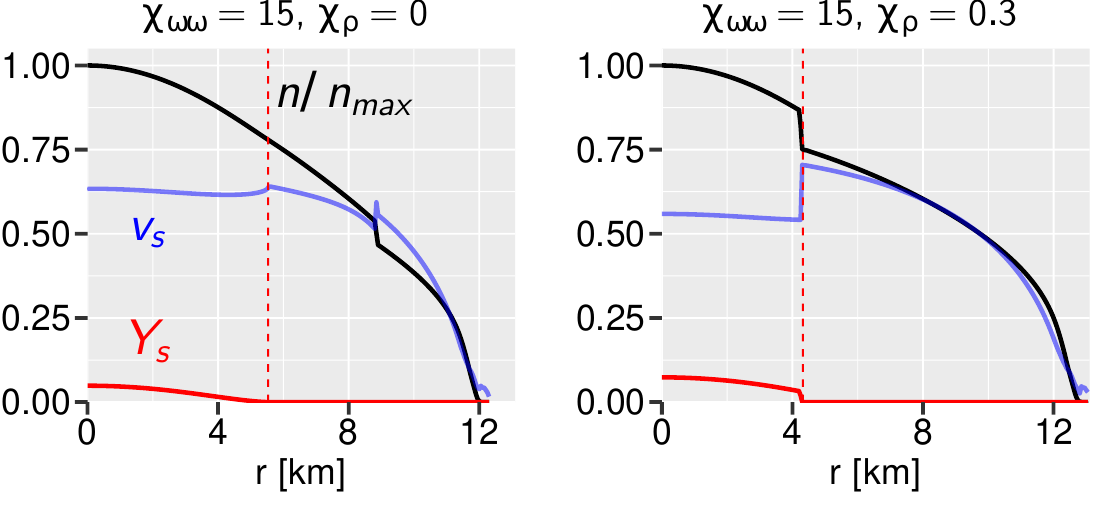}
\end{tabular}
	\caption{NS profile: the normalized baryonic density ($n/n_{\text{max}}$), 
 sound velocity ($v_s$ [c]), and strangeness fraction ($Y_s$) as a function of the radius for a $1.9M_{\odot}$ hybrid star. }
	\label{fig:profile19star}
	\end{center}
\end{figure}

Next, we analyze the profile of a $1.9M_{\odot}$ NS for specific hybrid star models.
Figure \ref{fig:profile19star} shows the density, sound velocity, and strangeness fraction as a function of NS radius for $\{ \chi_{\rho}=15, \chi_{\omega \omega}=0 \}$ (left) and $\{ \chi_{\rho}=15, \chi_{\omega \omega}=0.3 \}$ (right). We see that as $\chi_{\rho}$ increases, the onset of strangeness occurs at lower radius, i.e., gives rise to smaller quark cores. 
Actually for the $\{ \chi_{\rho}=15, \chi_{\omega \omega}=0.3 \}$ model, there are strange quarks in the whole quark core, i.e., there is a phase transition from hadronic matter directly to quark matter with strangeness, while for $\{ \chi_{\rho}=15, \chi_{\omega \omega}=0 \}$ there is a fraction of the quark core consisting of only light quarks. We also see that onset of strangeness is reflected on the presence of  a small bump in the speed of sound.

\begin{figure}[!htb]
\begin{center}
\begin{tabular}{c}
\includegraphics[width=0.45\linewidth]{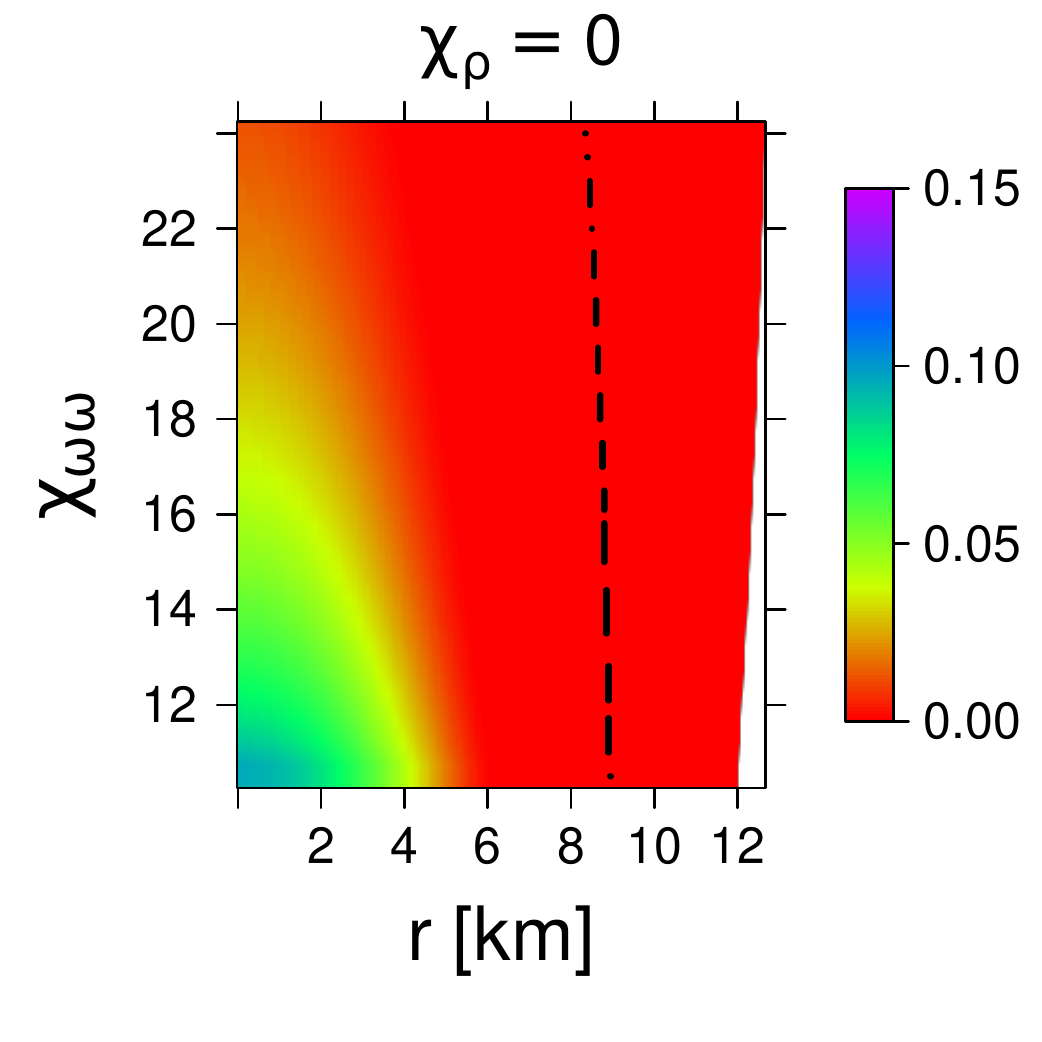}
\includegraphics[width=0.45\linewidth]{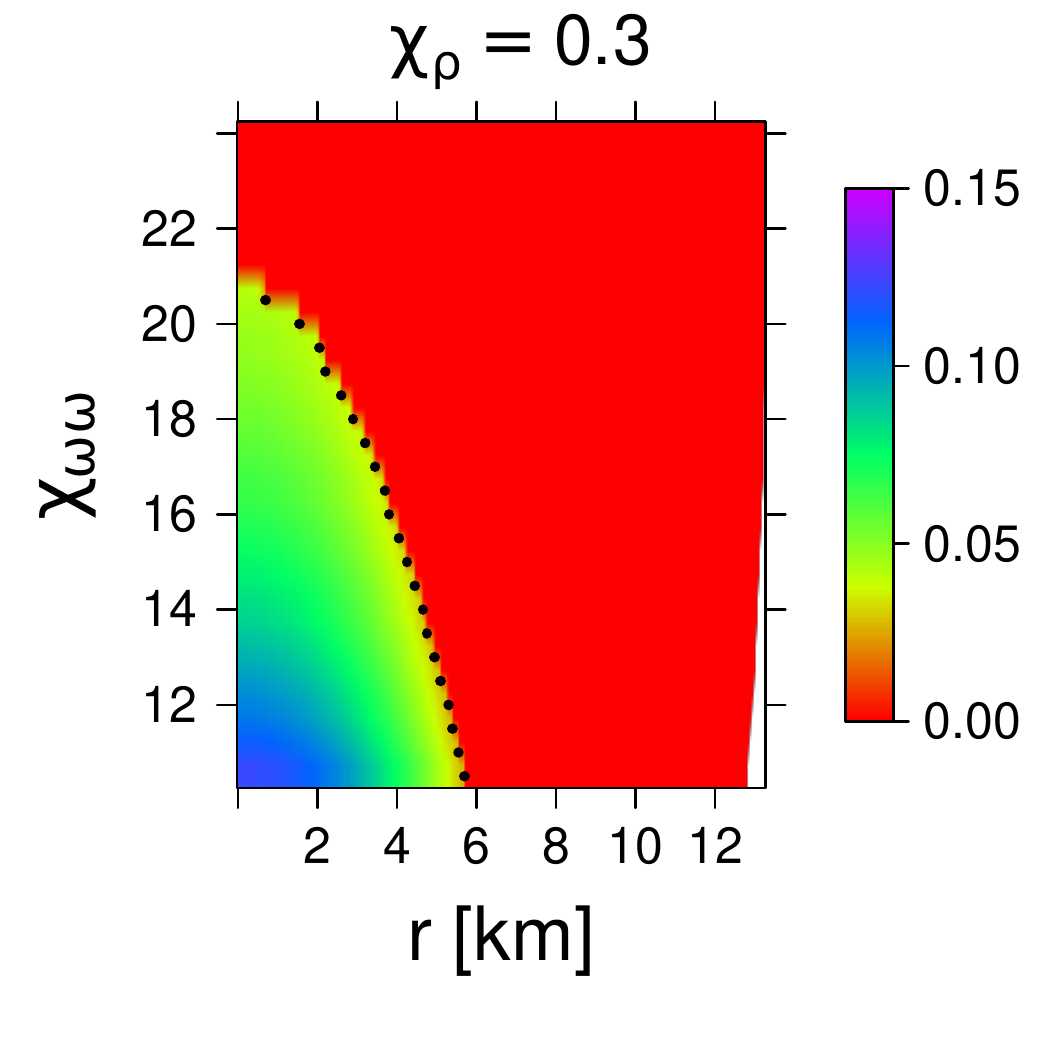}
\end{tabular}
	\caption{Strangeness profile for a $1.9M_{\odot}$ hybrid star as a function $\chi_{\omega \omega}$ for $\chi_{\rho}=0$ (left) and $\chi_{\rho}=0.3$ (right).}
	\label{fig:strangeness_2}
	\end{center}
\end{figure}

Figure \ref{fig:strangeness_2} shows how the strangeness content of a $1.9M_{\odot}$ hybrid star changes with $\{ \chi_{\rho}, \chi_{\omega \omega} \}$. For $\chi_{\rho}=0$ (left), 
we see that the phase transition from hadronic matter to quark matter (dashed line) occurs to a 2-flavor quark phase ($Y_s=0$) and the strangeness only appears for  $r<r_{hq}$, where $r_{hq}$ is the radius of the shell where the hadron-quark transition occurs. In other words, the quark core is made of an outer shell composed of light quarks only and a inner shell composed of all three quarks.
Other possible scenario is the one realized for $\chi_{\rho}=0.3$ (right), in which the quark core is totally made of the three quark flavours. Furthermore, there is no quark core for $\chi_{\omega \omega} > 21$ in a $1.9M_{\odot}$ NS.

\begin{figure}[!htb]
\begin{center}
\begin{tabular}{c}
\includegraphics[width=0.95\linewidth]{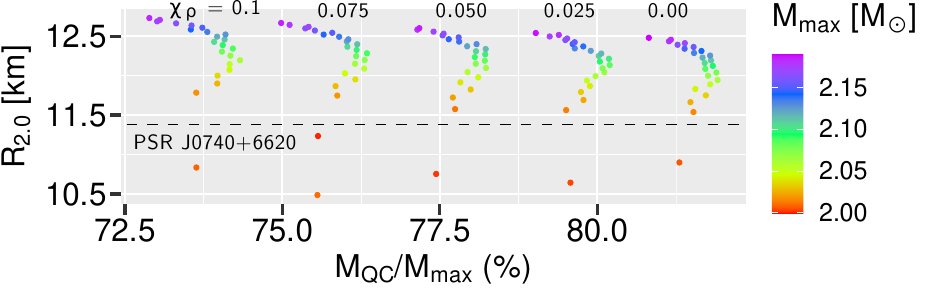}
\end{tabular}
	\caption{Radius of $2.0M_{\odot}$ hybrid stars as a function of the quark core mass fraction of the $M_{\text{max}}$. All models presented here predict a quark core in a $1.4M_{\odot}$ NS.}
	\label{fig:predictions}
	\end{center}
\end{figure}

Figure \ref{fig:predictions}  answers the following question: is it possible to have a quark core in a $1.4M_{\odot}$ NS and the hybrid stars within the same model still fulfil all available NS constrains? The figure shows the radius of $2.0M_{\odot}$ hybrid stars as a function of the fraction of the quark core mass with respect to the total star mass. We show hybrid star models with $\chi_{\rho}\le 0.1$ and $\chi_{\omega \omega}\le24$; however, instead of indicating the value of $\chi_{\omega \omega}$, we show the $M_{\text{max}}$ reached by each model ($M_{\text{max}}$ is proportional to $\chi_{\omega \omega}$). 
The plot also indicates the lower bond of the radius of PSR J0740+6620 (${2.072}_{-0.066}^{+0.067}M_{\odot}$) with $R={12.39}_{-0.98}^{+1.30}$ km  \cite{Riley:2021pdl}.
We conclude that almost all hybrid models  are able to describe the PSR J0740+6620, allowing for the quark core to vary from 71\% to 82\% of the $M_{\text{max}}$. A decrease of $\chi_{\rho}$ has the effect of increasing the quark core mass.

\section{Conclusions}
\label{sec:conclusions}

We have explored the effect of eight-quark vector and iso-vector interactions on the stability of hybrid stars. We show that the repulsive interaction is able to generate a quark core in $1.4M_{\odot}$ NS and still is able to reach $M_{\text{max}}\approx2.2 M_{\odot}$. 
The analysis of the GW170817 event indicates a low tidal deformability $\Lambda<700$ for a $1.4M_{\odot}$ NS, and thus hinting for lower NSs radii. However, the existence of massive NS favors the opposite limit, i.e., larger NS radii. We demonstrate that the hybrid star scenario, where the quark matter is realized inside NSs via a first-order phase transition, is able to accommodate, simultaneously, the existence of small mass hybrid NSs and has enough repulsion to sustain massive NSs, compatible with observations.

\end{document}